\documentclass[aps,prb,twocolumn,floatfix,superscriptaddress]{revtex4-1}

\usepackage{amsmath,amssymb,mathrsfs,color,graphicx,newtxtext,newtxmath,times}
\usepackage[colorlinks={true}, citecolor={blue}, filecolor={blue}, linkcolor={blue}, urlcolor={blue}]{hyperref}

\usepackage{soul}
\usepackage{filecontents}
\newcommand{\fref}[1]{Fig.~\ref{fig:#1}}
\newcommand{\Fref}[1]{Figure~\ref{fig:#1}}
\newcommand{\eref}[1]{Eq.~(\ref{eq:#1})}
\newcommand{\Eref}[1]{Equation~(\ref{eq:#1})}

\newcommand{\rmd}{\mathrm{d}}

\begin{document}

\title{Nonreciprocal topological phononics in optomechanical arrays}
\author{Claudio Sanavio}
\affiliation{Department of Physics, University of Malta, Msida MSD 2080, Malta}
\author{Vittorio Peano}
\affiliation{Department of Physics, University of Malta, Msida MSD 2080, Malta}
\affiliation{Max Planck Institute for the Science of Light, Staudstra{\ss}e 2, 91058 Erlangen, Germany}
\author{Andr\'e Xuereb}
\affiliation{Department of Physics, University of Malta, Msida MSD 2080, Malta}

\date{\today}

\begin{abstract}
We propose a platform for robust and tunable non-reciprocal phonon transport  based on arrays of optomechanical microtoroids. In our approach, time-reversal symmetry is broken by the interplay of photonic spin--orbit coupling, engineered using a state-of-the-art geometrical approach, and the optomechanical interaction. We demonstrate the topologically protected nature of this system by investigating its robustness to imperfections. This type of system could find application in phonon-based information storage and signal processing devices.
\end{abstract}

\maketitle

\section{Introduction}\label{I}

Tremendous progress in the control of radiation pressure in optomechanical systems has led to an impressive series of milestones including cooling of nanomechanical resonators to their quantum ground state, quantum-limited position measurements, and squeezing of quantum fluctuations~\cite{aspelmeyer_cavity_2014}. While the tunable interaction between a single optical and a single mechanical mode underlies most of the early breakthroughs, a new trend is emerging that exploits several optical and mechanical modes to perform more complex tasks. These include frequency conversion~\cite{andrews_bidirectional_2014}, robust synchronization~\cite{zhang_synchronization_2012}, and on-chip coherent nonreciprocal routing of photons~\cite{bernier_nonreciprocal_2016,fang_generalized_2017,peterson_demonstration_2017,xu_nonreciprocal_2018} and phonons~\cite{barzanjeh_mechanical_2017}. State-of-the-art nonreciprocal optomechanical devices are based on the interaction of a handful of modes. Since most optomechanical platforms are based on scalable on-chip architectures, a natural step forward will be to build devices based on periodic arrangements of colocalized optical and mechanical modes, so-called optomechanical arrays~\cite{heinrich_collective_2011,holmes_synchronization_2012,xuereb_strong_2012,schmidt_optomechanical_2012,ludwig_quantum_2013,xuereb_reconfigurable_2014,lauter_pattern_2015,schmidt_optomechanical_2015,schmidt_optomechanical_2015-1,zhang_synchronization_2015,peano_topological_2015,gan_solitons_2016}. An appealing advantage of this multimode approach compared to state-of-the-art few-mode nonreciprocal devices is that it promises to unleash topological protection against disorder~\cite{schmidt_optomechanical_2015-1,peano_topological_2015,mathew_synthetic_2018}. 

Here we show how to implement topologically robust nonreciprocal phonon transport in an array of optomechanical microtoroids. This is a tunable platform that has already been used to successfully demonstrate optically mediated mechanical synchronization~\cite{zhang_synchronization_2015}. In each microtoroid, a mechanical breathing mode is naturally coupled to whispering gallery optical modes of both clockwise and anticlockwise chirality. The bare nonlinear optomechanical coupling typically is very small; adding a laser drive leads to a much stronger and tuneable linear coupling~\cite{aspelmeyer_cavity_2014}. Following the approach of Ref.~\cite{hafezi_optomechanically_2012}, we use this  optomechanical tunability to select the chirality of the optical modes that are strongly coupled to the mechanical modes. Inspired by present-day implementations of optomechanical microtoroids arrays~\cite{zhang_synchronization_2015}, we assume that the optical driving is applied only at the edge of the device (see \fref{Kagome}). The photons hopping between neighboring toroids experience an effective spin--orbit coupling. This can be implemented following a well-tested approach based on asymmetric couplers~\cite{hafezi_robust_2011,hafezi_imaging_2013}. The interplay of the optical spin--orbit coupling and the laser driving induces a breaking of the time-reversal symmetry and leads to a mechanical Chern insulator supporting chiral phononic edge states. In stark contrast to geometry-based proposals~\cite{mousavi_topologically_2015,brendel_pseudomagnetic_2017,brendel_snowflake_2018} and experiments~\cite{yu_elastic_2018,cha_experimental_2018} for on-chip topological  phononics, our setup is truly nonreciprocal, and the topological protection extends to any arbitrary fabrication imperfection. Our approach differs from earlier proposals for on-chip mechanical Chern insulators~\cite{peano_topological_2015, mathew_synthetic_2018},  in that it does not require a direct mechanical coupling between the microtoroids and the driving field need not be applied to the bulk of the array.

Next, we introduce an optomechanical realization of a kagome lattice, and motivate a specific choice for the driving field. We then discuss the band structure of the system and the appearance of topologically protected mechanical edge states. Finally, we explore the transport properties of the system and demonstrate its robustness against disorder.
\begin{figure}
\includegraphics[width=1.\linewidth]{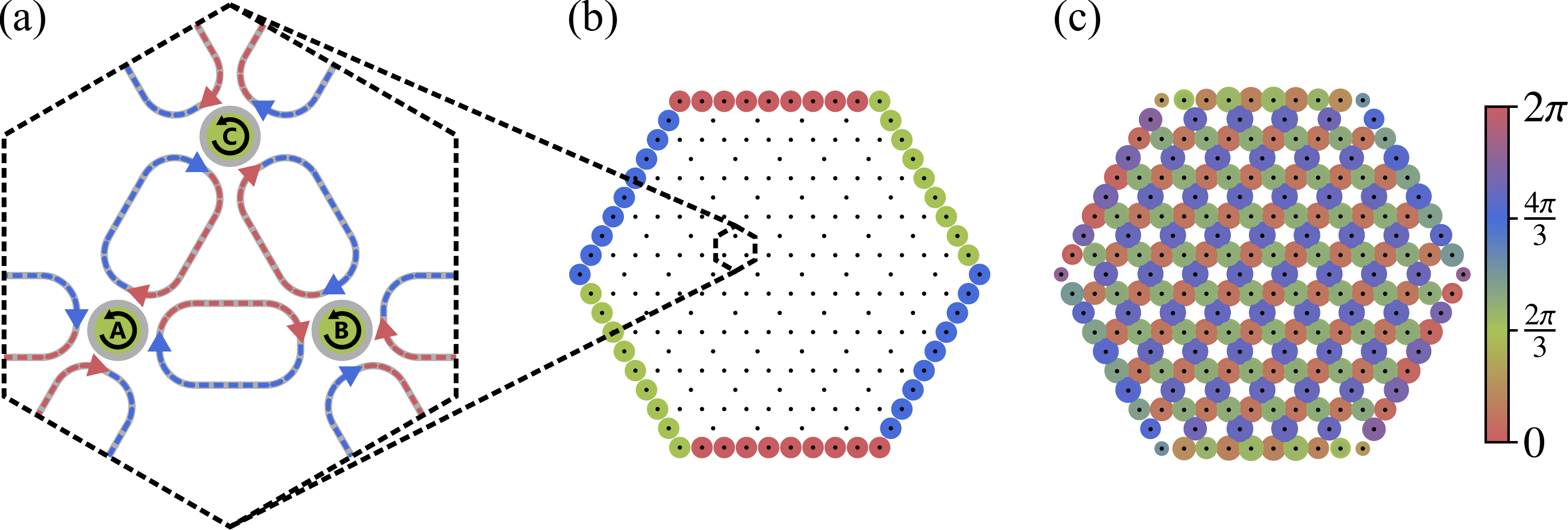}
\caption{(Color online) An optomechanical kagome lattice. (a)~The unit cell of the kagome lattice is composed of three optomechanical resonators linked by  off-center waveguides. (b)~We consider a finite hexagonal array of resonators driven optically at the edges. The driving fields have the same amplitude but different phases, as illustrated. This gives rise to (c)~a stationary field where the optical fields in adjacent sublattices have a phase difference of $2\pi/3$.}
\label{fig:Kagome}
\end{figure}

\begin{figure*}
\begin{center}
\includegraphics[width=\textwidth]{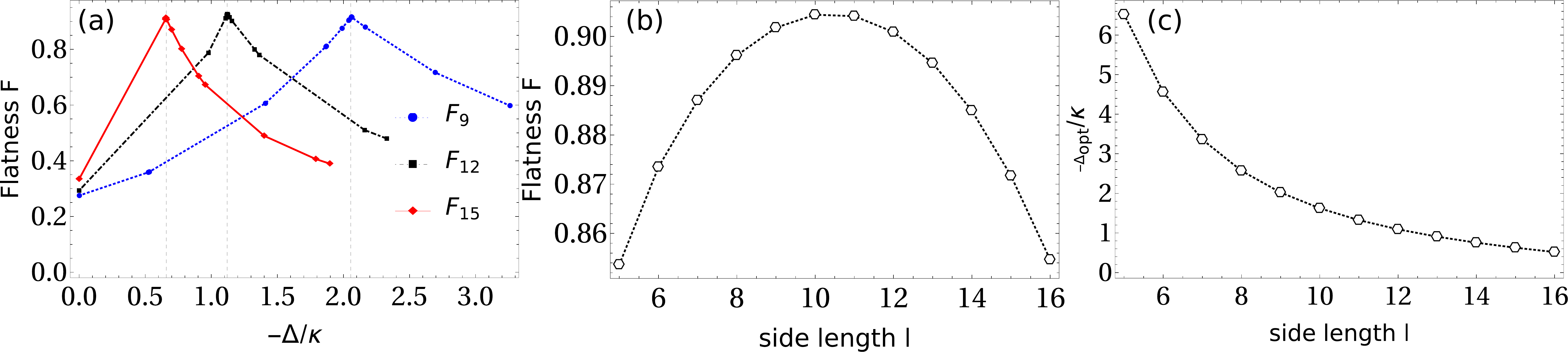}
\caption{(a)~ The parameter $F$ vs the detuning $\Delta$ in units of $\kappa$ for different side length $l$ of the hexagonal array. (b)~ The best values of $F$ for different sides' lengths. (c)~ The best detuning $\Delta_{opt}/\kappa$ to reach the highest value of $F$ as a function of the size $l$ of the array.}
\label{fig:FlatDelta}
\end{center}
\end{figure*}


\begin{figure}[h]
\includegraphics[width=.45\textwidth]{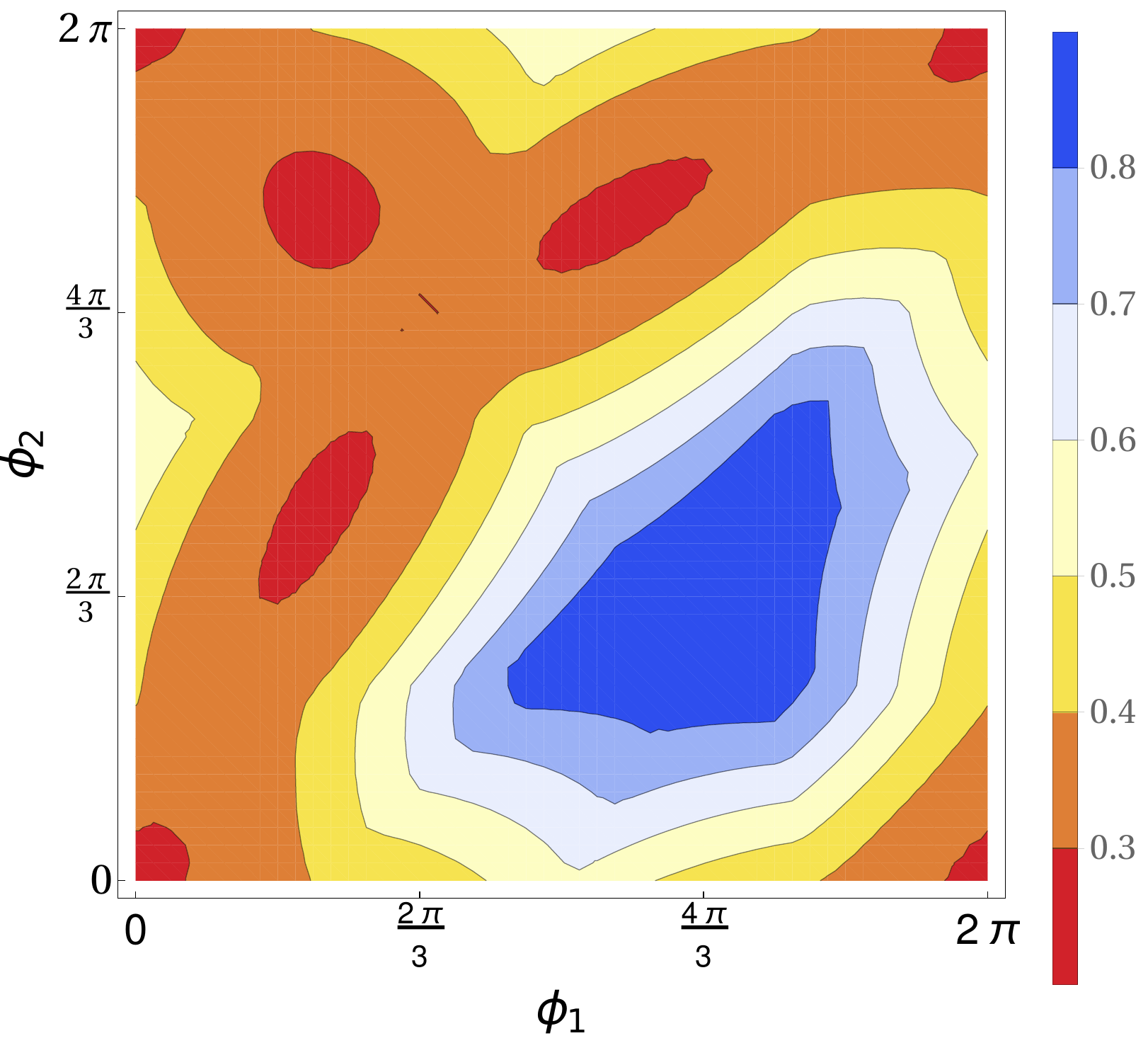}
\caption{$F$ as obtained varying the relative phases of the laser. The optimal conditions is obtained varying of $2\pi/3$ the phases of the lasers in clockwise direction.}
\label{fig:FlatPhase2d}
\end{figure}

\section{Optomechanical Kagome lattice}\label{II}

Our basic building block will be an optomechanical microtoroid resonator~\cite{aspelmeyer_cavity_2014}. These are well known to support degenerate clockwise and anticlockwise  optical modes coupled via radiation pressure to mechanical breathing modes. We shall be concerned with a kagome array of such resonators, \fref{Kagome}, each of which is described by the Hamiltonian
\begin{multline}
\label{eq:OMHamiltonian}
\hat{H}_{\mathbf{i}}=\sum_{\sigma}\bigg{(} \omega_{0}\hat{a}_{{\mathbf i},\sigma}^\dagger\hat{a}_{{\mathbf i},\sigma}+\Omega\hat{b}_{\mathbf i}^\dagger\hat{b}_{\mathbf i}+g_0\hat{a}_{{\mathbf i},\sigma}^\dagger\hat{a}_{{\mathbf i},\sigma}(\hat{b}_{\mathbf i}^\dagger+\hat{b}_{\mathbf i})\bigg{)}.
\end{multline}
Here, $g_0$ is the bare optomechanical coupling constant, while $\omega_0$ and $\Omega$ are the resonance frequencies of the optical and mechanical modes, respectively. The operator $\hat{b}_{\mathbf{i}}$ annihilates a phonon on site ${\mathbf i}=(i_1,i_2,i_\text{s})$, where $i_1,i_2 \in \mathbb{Z}$ identifies the unit cell and $i_\text{s}=\text{A},\text{B},\text{C}$ the sublattice. Likewise, $\hat{a}_{\mathbf{i},\sigma}$ are the annihilation operators for anticlockwise ($\sigma= +1$) and clockwise ($\sigma=-1$) photons. Decay of the optical and mechanical modes with the rates $\kappa$ and $\gamma$, respectively, is described by the standard input--output formalism~\cite{Gardiner_1985}. Throughout this paper, we will consider $\hbar=1$ for convenience.

In our setup, pairs of resonators are coupled evanescently via waveguide couplers, \fref{Kagome}(a). A photon propagating back and forth between two neighboring resonators through the coupler maintains its chirality $\sigma$. Moreover, on each leg of the round trip, it follows a different branch of the coupler [\fref{Kagome}(a)]. The length of the couplers is such that light of frequency $\sim\omega_\mathbf{i}$ interferes destructively inside them and they thus mediate a coherent nearest-neighbor interaction $\hat{H}_{\mathbf{ij}}$. This leads to the Hamiltonian $\hat{H}=\textstyle\sum_\mathbf{i}\hat{H}_\mathbf{i}+\textstyle\sum_{\langle \mathbf{i},\mathbf{j} \rangle}\hat{H}_{\mathbf{ij}}$, where  $\langle \mathbf{i},\mathbf{j} \rangle$ indicates nearest neighbors. Asymmetric couplers give rise to a spin--orbit-type interaction, $\hat{H}_{\mathbf{ij}}=\sum_\sigma \hat{H}_{\mathbf{ij},\sigma}$, where
\begin{equation}
\label{eq:HoppingHamiltonian}
\hat{H}_{\mathbf{ij},\sigma}=-J_{\mathbf{ij}}\bigl(e^{i\sigma\phi_{\mathbf{ij}}}\hat{a}_{\mathbf{i},\sigma}^\dagger\hat{a}_{\mathbf{j},\sigma}+e^{-i\sigma\phi_{\mathbf{ij}}}\hat{a}_{\mathbf{i},\sigma}\hat{a}_{\mathbf{j},\sigma}^\dagger\bigr).
\end{equation}
Here, $J_{\mathbf{ij}}$ is the hopping rate. For symmetry reasons, we consider $J_{\mathbf{ij}}\equiv{J}$. The complex phase $\phi_{\mathbf{ij}}$ is proportional to the length difference between the two branches of the coupler connecting resonators $\mathbf{i}$ and $\mathbf{j}$; a full derivation may be found in Ref.~\cite{hafezi_robust_2011}. We note that $\phi_{\mathbf{ij}}=-\phi_{\mathbf{ji}}$, such that $\hat{H}_{\mathbf{ij}}=\hat{H}_{\mathbf{ji}}$. Moreover, a difference in length between the internal and the external branches leads to a complex phase $\phi_{\mathbf{ij}}=\sigma\Phi/3$, where $\sigma$ is the chirality of the photon. After an anticlockwise circuit around a triangular plaquette, a photon picks up a phase $\sigma\Phi$.

\begin{figure*}
\begin{center}
\includegraphics[width=\linewidth]{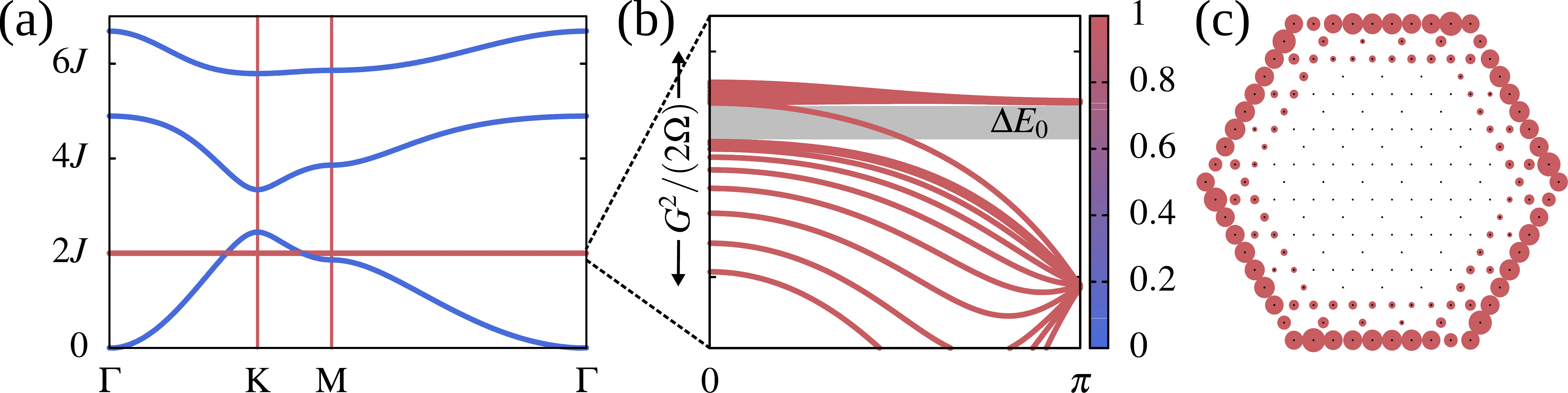}
\caption{(Color online) Band structure and the edge state. (a)~Bogoliubov--de Gennes particle--hole energies of the bulk Hamiltonian, \eref{LinearOMHamiltonian}, with quasi-momentum varying around the irreducible Brillouin zone of the kagome lattice, passing through the symmetry points $\Gamma$, $\text{K}$, and $\text{M}$. The color indicates the nature of the corresponding state, ranging from fully optical (blue) to fully mechanical (red). (b)~Dispersion relation of the infinite strip in the mechanical band. We find a state in the region labeled $\Delta E_0$ where, in the periodic case, there was a band gap. The energy $G^2/\Omega$ is indicated to set the scale. (c)~The state found in panel (b) is mechanical and confined to the edge of the finite array. Here, $G\sim0.005J$, $\Omega=2J$, and $\Delta E_0\sim2\times10^{-6}J$.}
\label{fig:BdG}
\end{center}
\end{figure*}

\section{Driving a finite size array}\label{IV}

\begin{figure}
\includegraphics[width=\linewidth]{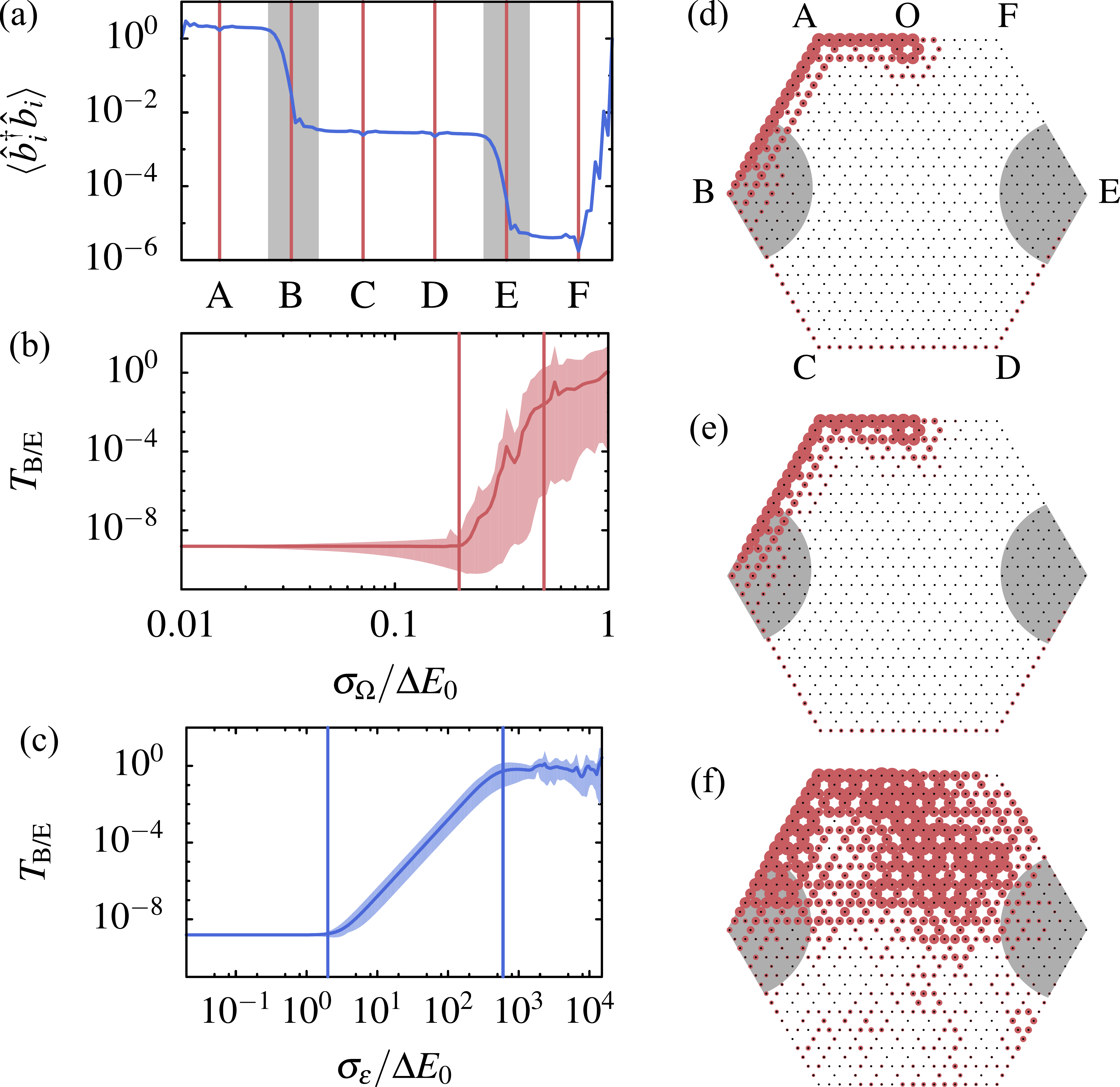}
\caption{(Color online) Topological transport. (a)~Magnitude of the internal field on the borders of the lattice when phonons are injected from the site labeled O in panel (d); the labels A and B also refer to panel (d). The field diminishes greatly at the sinks, represented by the gray areas around points B and E. (b)~Ratio between the mechanical output fields at points B and E, as a function of increasing mechanical disorder. For values of the variance of the disorder until about $0.2\Delta E_0$, the edge state and transport are not affected greatly. For increasing disorder, we find a phase transition and the ratio between the two output fields rapidly climbs to around $1$; there is no longer a preferred direction of transport. Each point in this plot is an average of $300$ realizations, with the shaded region indicating the range of values obtained. (c)~ Similar to panel (b) but as a function of increasing optical scattering. The edge state has a mostly mechanical character and is thus not greatly affected by the optical disorder until it increases to $\sim\Delta E_0$. Upon increasing disorder further to $\sim10^3\Delta E_0$, the system undergoes a topological phase transition and the directionality of the transport is lost. (d)--(f)~Realizations of the internal field where the variance of the mechanical disorder equals $0$, $0.1\Delta E_0$, and $0.5\Delta E_0$. Only in the latter figure the disorder is large enough to change the topological properties of the system.}
\label{fig:Transport}
\end{figure}

In our numerical studies, we consider a hexagonally shaped array with each side formed by $10$ resonators. To keep our arguments general, we will, however, consider an array with a total of $N$ sites. We drive the system optically through waveguides coupled evanescently to the resonators at the edge of the hexagonal array [\fref{Kagome}(c)]. Such a driving field is described by the Hamiltonian term
\begin{equation}
 \hat{H}_{\text{dr}}=i\sqrt{\kappa}\sum_{\mathbf{i}\ \in\ {\rm edge}}(f_{\mathbf{i}}e^{-i\omega_L t}\hat{a}_{\mathbf{i},+1}^\dagger-f^*_{\mathbf{i}}e^{i\omega_L t}\hat{a}_{\mathbf{i},+1}).
 \end{equation}
This driving field is chosen to have the same magnitude all around the edge. Pairs of opposite edges are driven at the same phase, while a phase difference of $2\pi/3$ is applied between pairs of adjacent edges, increasing in the clockwise orientation. In the interior of the lattice, as shown in \fref{Kagome}(c), this yields optical fields with magnitudes $\lvert\alpha_\mathbf{i}\rvert$ that depend only weakly on the site index $\mathbf{i}$ and whose phases are $0$ on sublattice A, $2\pi/3$ on sublattice B, and $4\pi/3$ on sublattice C. As a figure of merit for the quality of our solution, we define
\begin{equation}\label{flatness}
F=\Bigl(\sum_\mathbf{i}\lvert\alpha_\mathbf{i}\rvert\Bigr)\Big/\bigl(N\max_\mathbf{i}\{\lvert\alpha_\mathbf{i}\rvert\}\bigr),
\end{equation}
which ranges from $F=1/N$ for a situation where only one node has $\alpha_\mathbf{i}\neq0$, to $F=1$ when all the nodes have the same $\lvert\alpha_\mathbf{i}\rvert$.

To guarantee the stability of the dynamics, the array is red detuned and the analysis of $F$ has been restricted to values $\omega_l<\omega_0,$ where $\omega_0$ is the lowest eigenvalue of the system. 
Simulations show that it exists an optimal laser frequency $\omega_l$ that maximizes the parameter $F$. We find the optimal value of $\Delta=\omega_l-\omega_0$ depends on the size of the hexagonal array, and it decreases with bigger arrays, suggesting that for an infinite array the optimal detuning is at the lowest eigenstate frequency; see Fig.~\ref{fig:FlatDelta}(c). Although the optimal detuning $\Delta$ always exists, we find that the flatness $F$ is reduced for increasing size. This prevents us from applying this technique on large arrays, but it is worth noticing that current technology only allows a small number of microresonators in optomechanical arrays~\cite{zhang_synchronization_2015}.

Changing the relative phases among the driving field leads to an even better value for $F$. We found the optimal configuration built such that driving field of opposite sides have the same phases and the relative phases as shown in Fig. \ref{fig:FlatPhase2d}. In the work reported here, we have $F\geq0.9$.

\section{Time-reversal symmetry breaking}\label{III}

Chern insulators are two-dimensional materials marked by the presence of states localized at the edges. These states are topologically protected, meaning that their existence is guaranteed even if the system is subject to strong disorder as long as the topology of the bulk Hamiltonian is preserved. This means that the band structure can be modified by the disorder as long as it does not open nor close the band gap. When this happens, the system undergoes a topological phase transition. 
A precondition to create a Chern insulator is to break time-reversal symmetry. In our system, this is achieved by driving the array to a steady state with a large number of chirality-polarized circulating photons; implementation details for a finite-size array are discussed below. For concreteness, we consider the case where the circulating photons have anticlockwise chirality ($\alpha_{\mathbf{i}}\equiv\langle\hat{a}_{\mathbf{i},\sigma=1}\rangle_{\rm st}\gg 1$, $\langle\hat{a}_{\mathbf{i},\sigma=-1}\rangle_{\rm st}\approx 0$). In this regime, the clockwise photon modes are decoupled from the dynamics, since the optomechanical interaction is negligible at the single-photon level. A standard derivation~\cite{aspelmeyer_cavity_2014}  leads to a linearized Hamiltonian for the remaining degrees of freedom: $\hat{H}=\sum_{\mathbf{i}}\hat{H}_{\mathbf{i}}+\sum_{\langle \mathbf{i},\mathbf{j} \rangle}\hat{H}_{\mathbf{ij},1}$, where $\hat{H}_{\mathbf{ij},1}$ is given by \eref{HoppingHamiltonian} and
\begin{equation}
\label{eq:LinearOMHamiltonian}
\hat{H}_{\mathbf{i}}=-\Delta_\mathbf{i}\hat{a}_{\mathbf{i},+1}^\dagger\hat{a}_{\mathbf{i},+1}+\Omega\hat{b}_{\mathbf{i}}^\dagger\hat{b}_{\mathbf{i}}+(G_{\mathbf{i}}\hat{a}_{\mathbf{i},+1}^\dagger+G^*_{\mathbf{i}}\hat{a}_{\mathbf{i},+1})(\hat{b}_{\mathbf{i}}^\dagger+\hat{b}_{\mathbf{i}}).
\end{equation}
We note that this Hamiltonian is defined in a frame rotating with the laser frequency $\omega_L$, and the fields $\hat{a}_{\mathbf{i},+1}$ and $\hat{b}_{\mathbf{i}}$ now refer to deviations from their steady-state mean values  
$\alpha_{\mathbf{i}}=\langle \hat{a}_{\mathbf{i},+1}\rangle_{\rm st}$ 
and $\beta_{\mathbf{i}}=\langle \hat{b}_{\mathbf{i}}\rangle_{\rm st}$. We have also defined the detuning $\Delta=\omega_\text{L}-\omega_0$ and the  optomechanical coupling $G_{\mathbf{i}}=\alpha_{\mathbf{i}}g_0$.

We note that $G_\mathbf{i}$ may be complex since the phase of the light field may differ between sites. In the presence of a direct mechanical hopping term, this nontrivial pattern of phases in the optomechanical coupling can by itself break the time-reversal symmetry even in the absence of photonic spin--orbit coupling~\cite{peano_topological_2015}. However, in our system, this can be achieved only with a photonic spin--orbit coupling ($\Phi\neq0$).

\section{Mechanical edge states}\label{V}
A convenient means to understand the nature of the topological states borne by the system is to move to reciprocal space. To do this, we must cast the problem into a translationally invariant form. First, we assume that the parameters of the lattice nodes, i.e., $\Delta_\mathbf{i}$, $G_\mathbf{i}$, etc., are all identical, and we drop the index $i$; second, we add periodic boundary conditions to the system. \Eref{LinearOMHamiltonian} can then be written as a function of the quasimomentum in the two directions, $k_x$ and $k_y$. Before continuing, let us remark that the \eref{LinearOMHamiltonian} does not preserve the total number of excitations $\sum_\mathbf{i}(\hat{a}_\mathbf{i}^\dagger\hat{a}_\mathbf{i}+\hat{b}_\mathbf{i}^\dagger\hat{b}_\mathbf{i})$. We therefore need to apply the Bogoliubov--de Gennes (BdG) formalism, where we consider a particle--hole description of the system. It is possible to use the bosonic Bogoliubov--de Gennes transformation on Hamiltonian~(\ref{eq:LinearOMHamiltonian}) to find the eigenvalues of the system. For simplicity, we convert the vector index $\mathbf{i}$ to a scalar $i=1,\dots,N$. Denote by $\hat{\mathbf{c}}$ the vector of the operators $\hat{\mathbf{c}}=(\hat{a}_1,\dots,\hat{a}_N,\hat{b}_1,\dots,\hat{b}_N)$, which allows us to write
\begin{equation}\label{Hbdg}
\hat{H}=\begin{pmatrix}
\hat{\mathbf{c}}^\dagger, \hat{\mathbf{c}}
\end{pmatrix}
H_{\text{BdG}}
\begin{pmatrix}
\hat{\mathbf{c}}\\
\hat{\mathbf{c}}^\dagger
\end{pmatrix},
\end{equation}
where $H_{\text{BdG}}$ is a square matrix. In this picture, the temporal evolution of the operators can be written in the form
\begin{equation}\label{temp}
\frac{\rmd}{\rmd t}
\begin{pmatrix}
\hat{\mathbf{c}}\\
\hat{\mathbf{c}}^\dagger
\end{pmatrix}
=\sigma_zH_{\text{BdG}}
\begin{pmatrix}
\hat{\mathbf{c}}\\
\hat{\mathbf{c}}^\dagger
\end{pmatrix},
\end{equation}
with $\sigma_z$ being a diagonal $4N\times 4N$ matrix with elements on the diagonal equal $1$ for the first $2N$ entries and $-1$ for the last $2N$. The eigenvalues of the matrix $\sigma_\text{z}H_{\text{BdG}}$ give the energies in the particle-hole description of the system~\cite{Nam_2015}. In \fref{BdG}(a) we show the dispersion relations in the BdG picture, showing the high-symmetry points which form the corners of the first irreducible Brillouin zone, $\Gamma=(k_x=0,k_y=0)$, $\text{K}=(2\pi/3,2\pi/3)$, and $\text{M}=(\pi,0)$.

We set the mechanical frequency to coincide with an optical band. This choice leads to a repulsion between the optical and mechanical energy levels, creating a mechanical band whose size is of the order of $G^2/\Omega$. Inside this mechanical band, the optical features of the system induce a band gap. The emergence of edge states requires reducing the system to an infinite strip, i.e., where the periodic boundary conditions are applied in one direction only. As illustrated in \fref{BdG}(b), we subsequently find a mechanical edge state in the mechanical band gap. Restoring the finite size of our system, the resulting mechanical edge state is as shown in \fref{BdG}(c). The fact that this edge state appears only when we lift the boundary conditions is typical of Chern insulators. Despite the fact that this edge state is entirely mechanical in nature, it is induced in the mechanical oscillators indirectly through the optomechanical interaction and optical coupling between the nodes; there is \emph{no direct mechanical coupling} between the microtoroids.

\begin{figure*}[t!]
\begin{center}
\includegraphics[width=\textwidth]{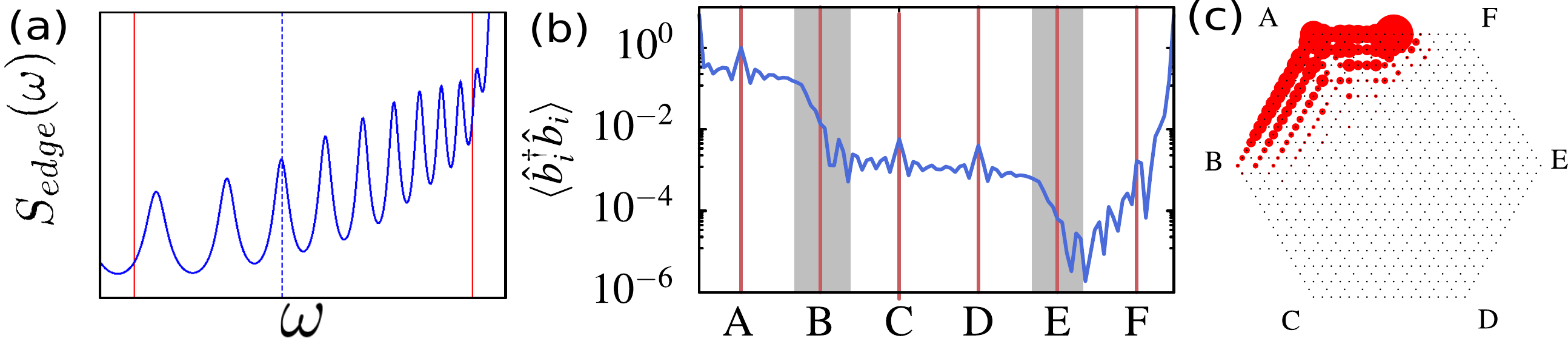}
\caption{(a)~The spectral density of a site on the edge in correspondence of the band gap $\Delta E$ delimited by the red lines and calculated with the values found in Ref. \cite{Verhagen2012}. The peak highlighted by the dotted line denotes an edge state whose transport behavior is shown in panels (b)~ and (c).}
\label{fig:edgestateV}
\end{center}
\end{figure*}
\section{Topologically protected transport}\label{VI}

We now turn our attention to the transport properties of the system. To do this we calculate the correlation between a phonon injected into site $\mathbf{j}$ at time $0$ transmitted to site $\mathbf{i}$ at time $t$. The Kubo formula $G_\text{MM}(t,\mathbf{i},\mathbf{j})=\langle\bigl[\hat{b}_\mathbf{i}(t),\hat{b}^\dagger_\mathbf{j}(0)\bigr]\rangle$ yields the transmission
\begin{equation}
T_\text{MM}(t,\mathbf{i},\mathbf{j})=\lvert G_\text{MM}(t,\mathbf{i},\mathbf{j})\rvert^2.
\end{equation}

The BdG description yields directly the matrices $\boldsymbol{G}(t)$ and $\boldsymbol{T}(t)$, which describe the correlations and transmission probabilities between all the creation and annihilation operators for all pairs of sites $\mathbf{i}$ and $\mathbf{j}$; i.e., the phonon--phonon elements of $\boldsymbol{G}(t)$ are $G_\text{MM}(t,\mathbf{i},\mathbf{j})$, etc., and similarly for $\boldsymbol{T}(t)$. It is convenient for our purposes to write $\boldsymbol{G}(t)$ in frequency space, $\boldsymbol{G}(\omega)=\int\rmd t e^{i\omega t}\boldsymbol{G}(t)$. We use this formula to calculate the internal field of a perturbation propagating inside the array.

Our starting point is a lattice of identical sites. To this we add two extended mechanical sinks centered around corners B and E of the hexagon, as shown in Figs. \ref{fig:Transport}(d)-\ref{fig:Transport}(f). To avoid perturbing the edge state, these sinks were introduced through a smooth additional mechanical decay rate $\gamma_{\text{ext},\mathbf{i}}$. The topological protection of the edge state means that if the function used were not smooth enough the edge state would simply avoid the source. In our case, we use the Fermi function
$\gamma_{\text{ext},\mathbf{i}}={2\gamma_{\text{ext},\text{max}}}\big/\bigl[{e^{d(\mathbf{i},\mathbf{j})/D}+1}\bigr],$
where $\mathbf{i}$ is the site in question, $\mathbf{j}$ is the site around which the sink is centered, i.e., the site at corner B or E in \fref{Transport}, $d(\mathbf{i},\mathbf{j})$ is the distance between the two sites, and $D$ sets the size of the sink. This decay rate changes smoothly from $\gamma_{\text{ext},\text{max}}$ at the center of the sink to zero when $d(\mathbf{i},\mathbf{j})\gg D$. In Figs. \ref{fig:Transport}(a) and \ref{fig:Transport}(d)--\ref{fig:Transport}(f) we shade the two regions for which $d(\mathbf{i},\mathbf{j})<D$ in gray.

In the following, we shall be concerned only with transport of phonons through the system. We drive the mechanical system at position O with a frequency resonant with the edge state identified in \fref{BdG}. As illustrated clearly in Figs. \ref{fig:Transport}(a) and \ref{fig:Transport}(d), the mechanical excitation travels along the edge state in the anti-clockwise direction. The transmission amplitude decreases slightly along the way because of the intrinsic optical and mechanical losses. Eventually, the excitation enters the region of the sink centered around corner B, whereupon it suddenly decreases in amplitude. The absorption of the sink is not perfect, and some of the mechanical excitation reaches the second sink, centered around corner E. The rapid increase in \fref{Transport}(a) at corner F is due to the limited, but finite, propagation of the mechanical excitation in the ``wrong,'' clockwise, orientation.

In \fref{Transport}(a), the slope of the curve is not zero away from the sinks. As mentioned earlier, this is due to the optical and mechanical losses inherent in the system, in conjunction with the finite velocity of the edge state. We verified this by comparing the observed slope $\Gamma_\text{obs}$, which is the same throughout the plot away from the shaded regions, to $\Gamma_\text{dir}$, derived directly from the band structure. To calculate $\Gamma_\text{dir}=\Gamma/v$ we divided the decay rate $\Gamma$ of the spectrum of the system, obtained from the Lorentzian curve found at the frequency in question, by the velocity $v$ of the edge state. This is calculated from the slope of the edge state at this frequency, and is derived from the dispersion relation shown in \fref{BdG}(b). We verified throughout this paper that $\Gamma_\text{obs}=\Gamma_\text{dir}$.

One of the most technologically important features of edge states of a Chern insulator is that they minimize the effects of back scattering. Whereas in ordinary conductors a wave that travels in one direction will be reflected off of a discontinuity, Chern insulators are immune to such scattering as long as the discontinuity leaves unchanged the topology of the dispersion relation. To verify this statement and illustrate its implications in our system, we inserted mechanical disorder in the system. This is described by a modification of the on-site Hamiltonian,
\begin{equation}
\hat{H}_\mathbf{i}\to\hat{H}_\mathbf{i}+\delta\Omega_\mathbf{i}\,\hat{b}_\mathbf{i}^\dagger\hat{b}_\mathbf{i},
\end{equation}
where $\delta\Omega_\mathbf{i}$ is drawn from a zero-mean Gaussian distribution with variance $\sigma_\Omega$. As we will show, the nature of the edge state and transport remain relatively unaffected for small $\sigma_\Omega$, until the disorder is large enough to close the band gap.

To quantify the level of robustness against disorder, we use the ratio between the absolute values of the output mechanical fields from the sinks centered at B and E. In turn, this is calculated using the input--output relation $\hat{b}_\mathbf{i}^\text{(out)}=\hat{b}_\mathbf{i}^\text{(in)}-\sqrt{\gamma_{\text{ext},\mathbf{i}}}\hat{b}_i$ for the mechanical fields; the relevant quantity is $T_{\text{B}/\text{E}}={\lvert\hat{b}_\text{B}^\text{(out)}\rvert^2}\big/{\lvert\hat{b}_\text{E}^\text{(out)}\rvert^2}.$
In \fref{Transport}(b), we illustrate this quantity as a function of increasing variance of the disorder. To ease interpretation, we scale the disorder by the scale of the band gap, $\Delta E_0$. We can see that for $\sigma_\Omega$ up to $\sim0.2\Delta E_0$ the disorder affects neither the direction of propagation of the edge state, as shown in \fref{Transport}(b), nor the nature of the edge state, as can be seen by comparing \fref{Transport}(e) to \fref{Transport}(d). For larger values of the disorder, we encounter a phase transition where the slope of the curve in \fref{Transport}(b) suddenly changes. This indicates a breakdown of topologically protected transport in the system and coincides with the edge state losing its nature entirely, cf.\ \fref{Transport}(f).

The two chiral optical modes can interact through scattering. This results in additional terms in the Hamiltonian,
\begin{equation}
\hat{H}_\mathbf{i}\to\hat{H}_\mathbf{i}+\varepsilon_\mathbf{i}\bigl(\hat{a}^\dagger_{\mathbf{i},+1}\hat{a}_{\mathbf{i},-1}+\hat{a}_{\mathbf{i},+1}\hat{a}^\dagger_{\mathbf{i},-1}\bigr),
\end{equation}
where $\varepsilon_\mathbf{i}$ is a random variable drawn from a zero-mean Gaussian distribution with variance $\sigma_\varepsilon$. As we did for the mechanical disorder, we show in \fref{Transport}(c) the ratio $T_{\text{B}/\text{E}}$ as a function of $\sigma_\varepsilon$. The system keeps its topological properties until $\sigma_\varepsilon\sim\Delta E_0$, beyond which it again undergoes a topological phase transition where the transport loses its directionality.

\section{Decay rate of the edge state and experimental realization}
\label{VIII}

In order to calculate the decay rate that affects the edge state, we analyze the spectral density matrix $S(\omega)$ of the system whose component are defined as

\begin{equation}\label{spectraldensity}
S_{ij}(\omega)=\int dt e^{i\omega t}\langle F_i(t)F_j(0)\rangle,
\end{equation}
\noindent where $F_i\sim g(a_i+a_i^{\dagger})$ is the linearized radiation pressure force acting on the site $i$. A simple calculation allows us to express $S(\omega)$ in terms of the Green's function $\boldsymbol{G}(\omega)$

\begin{equation}\label{spectraldensityarray}
S_{ij}(\omega)\sim \sum_{kl}\boldsymbol{G}_{ik}^\dagger \boldsymbol{G}_{lj},
\end{equation}

\noindent The resulting power spectrum is obtained looking at the diagonal elements of $S$. For a site $i$ belonging to the edge of the array, the function $S_{ii}(\omega)$ shows peaks in correspondence of the edge states; see \fref{edgestateV}, panel (a). The peaks can be fitted with a Lorentzian curve whose width is the decay rate $\Gamma$.

To better link our analysis to current experimental capabilities, we calculated the mechanical band gap $\Delta E$ and the decay rate $\Gamma$ with the parameters taken from Ref.~\cite{Verhagen2012} that regards a system composed by a single optomechanical microtoroid. 
\noindent To allow the comparison, we set the mechanical frequency $\Omega=2$ as in the main text and we kept unchanged the proportionality among the other values. The resulting frequencies are then in accordance with values sourced from the literature. We then extended the system to an array of such microresonators and introduced the optical hopping strength $J$ and performed the power spectrum analysis. Whenever the decay rate $\Gamma$ is smaller than the band gap, the transmission becomes possible. \Fref{BandGapVsJ} shows that in a range of $J$ between $0.15$ and $1$ the condition $\Gamma<\Delta E$ is satisfied. We notice that at the value $J/\Omega=1/2$ the ratio $J/\Delta E\sim3\times10^{-2}$.Figures \ref{fig:edgestateV}(b) and \ref{fig:edgestateV}(c) show the transport of the edge state for the experimental values of Ref.~\cite{Verhagen2012} in the case of zero disorder. The results are in agreement with what we found in our numerical analysis of the previous section.
\begin{figure}[t!]
\includegraphics[width=.45\textwidth]{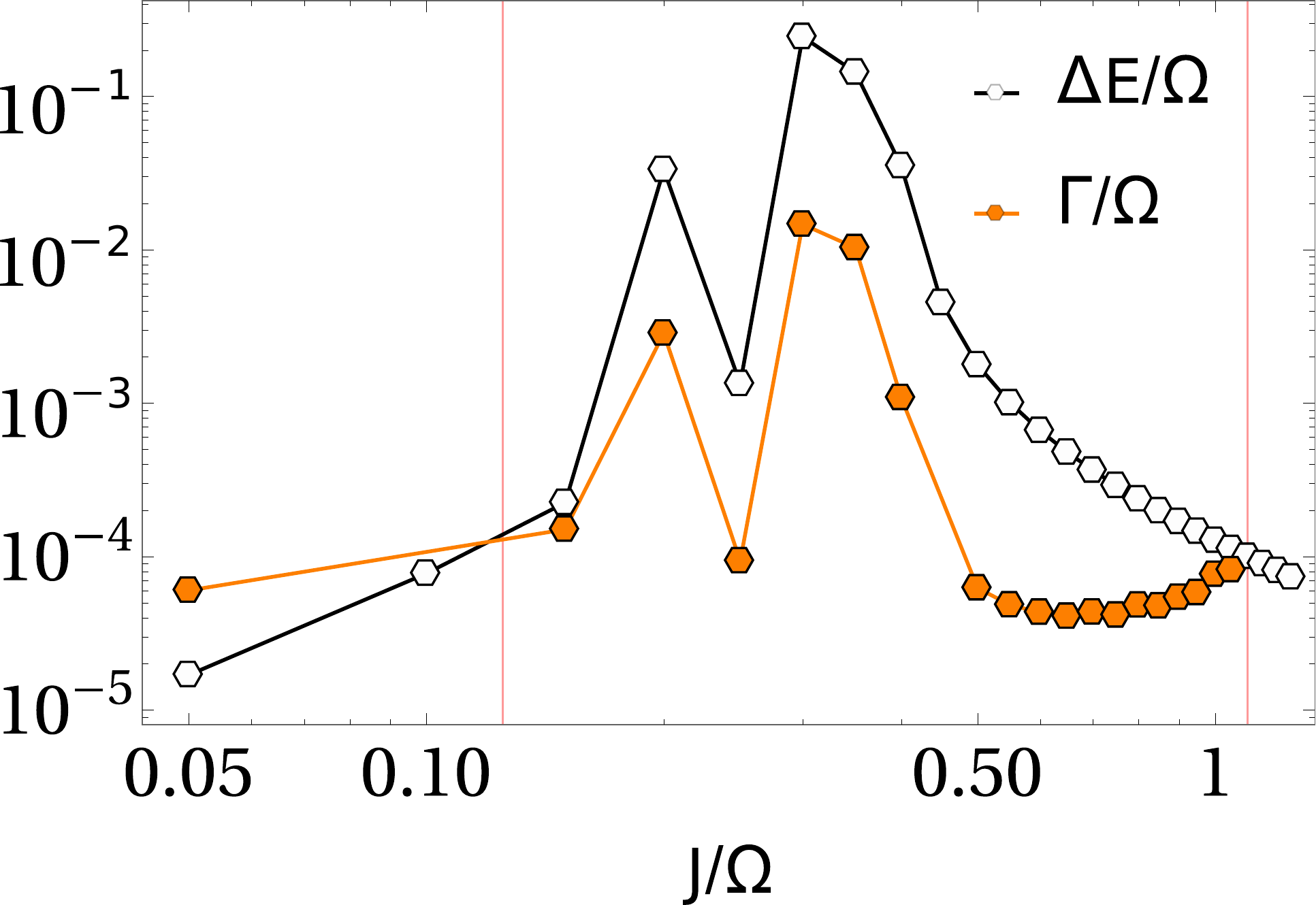}
\caption{Comparison between the mechanical band gap and the edge-state decay rate calculated from the power spectrum analysis, varying the optical hopping $J/\Omega$. Here $g/\Omega=0.025,\kappa/\Omega=0.009,\gamma/\Omega=5\times10^{-6}$. The numbers are chosen in order to be coherent with Ref. \cite{Verhagen2012}. The calculations in this work refer to a situation with $J/\Omega=1/2.$}
\label{fig:BandGapVsJ}
\end{figure}


\section{Conclusions}\label{VII}

We have described a system consisting of optomechanical microtoroids arranged in a kagome lattice. Optical coupling between the toroids gives rise to an effective spin--orbit interaction for the photons in the array. Through the optomechanical interaction and a judicious choice of the optical driving field, this gives rise to topologically protected mechanical edge states which are protected against fabrication and other disorder. Furthermore, we investigate the realizability of our model with experimentally suitable parameters. Our proposal can be realized using available technology and may be useful for phonon-based information storage and signal processing.

\section{Acknowledgments}

This work is supported by the European Union's Horizon 2020 for research and innovation programme under Grant Agreement No.\ 732894 (FET Proactive HOT). We acknowledge funding from the Julian Schwinger Foundation.

\bibliographystyle{apsrev4-1}

\bibliography{biblio}

\begin{thebibliography}{32}%
\makeatletter
\providecommand \@ifxundefined [1]{%
 \@ifx{#1\undefined}
}%
\providecommand \@ifnum [1]{%
 \ifnum #1\expandafter \@firstoftwo
 \else \expandafter \@secondoftwo
 \fi
}%
\providecommand \@ifx [1]{%
 \ifx #1\expandafter \@firstoftwo
 \else \expandafter \@secondoftwo
 \fi
}%
\providecommand \natexlab [1]{#1}%
\providecommand \enquote  [1]{``#1''}%
\providecommand \bibnamefont  [1]{#1}%
\providecommand \bibfnamefont [1]{#1}%
\providecommand \citenamefont [1]{#1}%
\providecommand \href@noop [0]{\@secondoftwo}%
\providecommand \href [0]{\begingroup \@sanitize@url \@href}%
\providecommand \@href[1]{\@@startlink{#1}\@@href}%
\providecommand \@@href[1]{\endgroup#1\@@endlink}%
\providecommand \@sanitize@url [0]{\catcode `\\12\catcode `\$12\catcode
  `\&12\catcode `\#12\catcode `\^12\catcode `\_12\catcode `\%12\relax}%
\providecommand \@@startlink[1]{}%
\providecommand \@@endlink[0]{}%
\providecommand \url  [0]{\begingroup\@sanitize@url \@url }%
\providecommand \@url [1]{\endgroup\@href {#1}{\urlprefix }}%
\providecommand \urlprefix  [0]{URL }%
\providecommand \Eprint [0]{\href }%
\providecommand \doibase [0]{http://dx.doi.org/}%
\providecommand \selectlanguage [0]{\@gobble}%
\providecommand \bibinfo  [0]{\@secondoftwo}%
\providecommand \bibfield  [0]{\@secondoftwo}%
\providecommand \translation [1]{[#1]}%
\providecommand \BibitemOpen [0]{}%
\providecommand \bibitemStop [0]{}%
\providecommand \bibitemNoStop [0]{.\EOS\space}%
\providecommand \EOS [0]{\spacefactor3000\relax}%
\providecommand \BibitemShut  [1]{\csname bibitem#1\endcsname}%
\let\auto@bib@innerbib\@empty
\bibitem [{\citenamefont {Aspelmeyer}\ \emph {et~al.}(2014)\citenamefont
  {Aspelmeyer}, \citenamefont {Kippenberg},\ and\ \citenamefont
  {Marquardt}}]{aspelmeyer_cavity_2014}%
  \BibitemOpen
  \bibfield  {author} {\bibinfo {author} {\bibfnamefont {M.}~\bibnamefont
  {Aspelmeyer}}, \bibinfo {author} {\bibfnamefont {T.~J.}\ \bibnamefont
  {Kippenberg}}, \ and\ \bibinfo {author} {\bibfnamefont {F.}~\bibnamefont
  {Marquardt}},\ }\href {\doibase 10.1103/RevModPhys.86.1391} {\bibfield
  {journal} {\bibinfo  {journal} {Reviews of Modern Physics}\ }\textbf
  {\bibinfo {volume} {86}},\ \bibinfo {pages} {1391} (\bibinfo {year}
  {2014})}\BibitemShut {NoStop}%
\bibitem [{\citenamefont {Andrews}\ \emph {et~al.}(2014)\citenamefont
  {Andrews}, \citenamefont {Peterson}, \citenamefont {Purdy}, \citenamefont
  {Cicak}, \citenamefont {Simmonds}, \citenamefont {Regal},\ and\ \citenamefont
  {Lehnert}}]{andrews_bidirectional_2014}%
  \BibitemOpen
  \bibfield  {author} {\bibinfo {author} {\bibfnamefont {R.~W.}\ \bibnamefont
  {Andrews}}, \bibinfo {author} {\bibfnamefont {R.~W.}\ \bibnamefont
  {Peterson}}, \bibinfo {author} {\bibfnamefont {T.~P.}\ \bibnamefont {Purdy}},
  \bibinfo {author} {\bibfnamefont {K.}~\bibnamefont {Cicak}}, \bibinfo
  {author} {\bibfnamefont {R.~W.}\ \bibnamefont {Simmonds}}, \bibinfo {author}
  {\bibfnamefont {C.~A.}\ \bibnamefont {Regal}}, \ and\ \bibinfo {author}
  {\bibfnamefont {K.~W.}\ \bibnamefont {Lehnert}},\ }\href
  {https://doi.org/10.1038/nphys2911} {\bibfield  {journal} {\bibinfo
  {journal} {Nature Physics}\ }\textbf {\bibinfo {volume} {10}},\ \bibinfo
  {pages} {321} (\bibinfo {year} {2014})}\BibitemShut {NoStop}%
\bibitem [{\citenamefont {Zhang}\ \emph {et~al.}(2012)\citenamefont {Zhang},
  \citenamefont {Wiederhecker}, \citenamefont {Manipatruni}, \citenamefont
  {Barnard}, \citenamefont {McEuen},\ and\ \citenamefont
  {Lipson}}]{zhang_synchronization_2012}%
  \BibitemOpen
  \bibfield  {author} {\bibinfo {author} {\bibfnamefont {M.}~\bibnamefont
  {Zhang}}, \bibinfo {author} {\bibfnamefont {G.~S.}\ \bibnamefont
  {Wiederhecker}}, \bibinfo {author} {\bibfnamefont {S.}~\bibnamefont
  {Manipatruni}}, \bibinfo {author} {\bibfnamefont {A.}~\bibnamefont
  {Barnard}}, \bibinfo {author} {\bibfnamefont {P.}~\bibnamefont {McEuen}}, \
  and\ \bibinfo {author} {\bibfnamefont {M.}~\bibnamefont {Lipson}},\ }\href
  {\doibase 10.1103/PhysRevLett.109.233906} {\bibfield  {journal} {\bibinfo
  {journal} {Physical Review Letters}\ }\textbf {\bibinfo {volume} {109}},\
  \bibinfo {pages} {233906} (\bibinfo {year} {2012})}\BibitemShut {NoStop}%
\bibitem [{\citenamefont {Bernier}\ \emph {et~al.}(2017)\citenamefont
  {Bernier}, \citenamefont {T{\'o}th}, \citenamefont {Koottandavida},
  \citenamefont {Ioannou}, \citenamefont {Malz}, \citenamefont {Nunnenkamp},
  \citenamefont {Feofanov},\ and\ \citenamefont
  {Kippenberg}}]{bernier_nonreciprocal_2016}%
  \BibitemOpen
  \bibfield  {author} {\bibinfo {author} {\bibfnamefont {N.~R.}\ \bibnamefont
  {Bernier}}, \bibinfo {author} {\bibfnamefont {L.~D.}\ \bibnamefont
  {T{\'o}th}}, \bibinfo {author} {\bibfnamefont {A.}~\bibnamefont
  {Koottandavida}}, \bibinfo {author} {\bibfnamefont {M.~A.}\ \bibnamefont
  {Ioannou}}, \bibinfo {author} {\bibfnamefont {D.}~\bibnamefont {Malz}},
  \bibinfo {author} {\bibfnamefont {A.}~\bibnamefont {Nunnenkamp}}, \bibinfo
  {author} {\bibfnamefont {A.~K.}\ \bibnamefont {Feofanov}}, \ and\ \bibinfo
  {author} {\bibfnamefont {T.~J.}\ \bibnamefont {Kippenberg}},\ }\href
  {\doibase 10.1038/s41467-017-00447-1} {\bibfield  {journal} {\bibinfo
  {journal} {Nature Communications}\ }\textbf {\bibinfo {volume} {8}},\
  \bibinfo {pages} {604} (\bibinfo {year} {2017})}\BibitemShut {NoStop}%
\bibitem [{\citenamefont {Fang}\ \emph {et~al.}(2017)\citenamefont {Fang},
  \citenamefont {Luo}, \citenamefont {Metelmann}, \citenamefont {Matheny},
  \citenamefont {Marquardt}, \citenamefont {Clerk},\ and\ \citenamefont
  {Painter}}]{fang_generalized_2017}%
  \BibitemOpen
  \bibfield  {author} {\bibinfo {author} {\bibfnamefont {K.}~\bibnamefont
  {Fang}}, \bibinfo {author} {\bibfnamefont {J.}~\bibnamefont {Luo}}, \bibinfo
  {author} {\bibfnamefont {A.}~\bibnamefont {Metelmann}}, \bibinfo {author}
  {\bibfnamefont {M.~H.}\ \bibnamefont {Matheny}}, \bibinfo {author}
  {\bibfnamefont {F.}~\bibnamefont {Marquardt}}, \bibinfo {author}
  {\bibfnamefont {A.~A.}\ \bibnamefont {Clerk}}, \ and\ \bibinfo {author}
  {\bibfnamefont {O.}~\bibnamefont {Painter}},\ }\href
  {https://doi.org/10.1038/nphys4009} {\bibfield  {journal} {\bibinfo
  {journal} {Nature Physics}\ }\textbf {\bibinfo {volume} {13}},\ \bibinfo
  {pages} {465} (\bibinfo {year} {2017})}\BibitemShut {NoStop}%
\bibitem [{\citenamefont {Peterson}\ \emph {et~al.}(2017)\citenamefont
  {Peterson}, \citenamefont {Lecocq}, \citenamefont {Cicak}, \citenamefont
  {Simmonds}, \citenamefont {Aumentado},\ and\ \citenamefont
  {Teufel}}]{peterson_demonstration_2017}%
  \BibitemOpen
  \bibfield  {author} {\bibinfo {author} {\bibfnamefont {G.~A.}\ \bibnamefont
  {Peterson}}, \bibinfo {author} {\bibfnamefont {F.}~\bibnamefont {Lecocq}},
  \bibinfo {author} {\bibfnamefont {K.}~\bibnamefont {Cicak}}, \bibinfo
  {author} {\bibfnamefont {R.~W.}\ \bibnamefont {Simmonds}}, \bibinfo {author}
  {\bibfnamefont {J.}~\bibnamefont {Aumentado}}, \ and\ \bibinfo {author}
  {\bibfnamefont {J.~D.}\ \bibnamefont {Teufel}},\ }\href {\doibase
  10.1103/PhysRevX.7.031001} {\bibfield  {journal} {\bibinfo  {journal} {Phys.
  Rev. X}\ }\textbf {\bibinfo {volume} {7}},\ \bibinfo {pages} {031001}
  (\bibinfo {year} {2017})}\BibitemShut {NoStop}%
\bibitem [{\citenamefont {Xu}\ \emph {et~al.}(2019)\citenamefont {Xu},
  \citenamefont {Jiang}, \citenamefont {Clerk},\ and\ \citenamefont
  {Harris}}]{xu_nonreciprocal_2018}%
  \BibitemOpen
  \bibfield  {author} {\bibinfo {author} {\bibfnamefont {H.}~\bibnamefont
  {Xu}}, \bibinfo {author} {\bibfnamefont {L.}~\bibnamefont {Jiang}}, \bibinfo
  {author} {\bibfnamefont {A.~A.}\ \bibnamefont {Clerk}}, \ and\ \bibinfo
  {author} {\bibfnamefont {J.~G.~E.}\ \bibnamefont {Harris}},\ }\href {\doibase
  10.1038/s41586-019-1061-2} {\bibfield  {journal} {\bibinfo  {journal} {Nature
  (London)}\ }\textbf {\bibinfo {volume} {568}},\ \bibinfo {pages} {65}
  (\bibinfo {year} {2019})}\BibitemShut {NoStop}%
\bibitem [{\citenamefont {Barzanjeh}\ \emph {et~al.}(2017)\citenamefont
  {Barzanjeh}, \citenamefont {Wulf}, \citenamefont {Peruzzo}, \citenamefont
  {Kalaee}, \citenamefont {Dieterle}, \citenamefont {Painter},\ and\
  \citenamefont {Fink}}]{barzanjeh_mechanical_2017}%
  \BibitemOpen
  \bibfield  {author} {\bibinfo {author} {\bibfnamefont {S.}~\bibnamefont
  {Barzanjeh}}, \bibinfo {author} {\bibfnamefont {M.}~\bibnamefont {Wulf}},
  \bibinfo {author} {\bibfnamefont {M.}~\bibnamefont {Peruzzo}}, \bibinfo
  {author} {\bibfnamefont {M.}~\bibnamefont {Kalaee}}, \bibinfo {author}
  {\bibfnamefont {P.~B.}\ \bibnamefont {Dieterle}}, \bibinfo {author}
  {\bibfnamefont {O.}~\bibnamefont {Painter}}, \ and\ \bibinfo {author}
  {\bibfnamefont {J.~M.}\ \bibnamefont {Fink}},\ }\href {\doibase
  10.1038/s41467-017-01304-x} {\bibfield  {journal} {\bibinfo  {journal}
  {Nature Communications}\ }\textbf {\bibinfo {volume} {8}},\ \bibinfo {pages}
  {953} (\bibinfo {year} {2017})}\BibitemShut {NoStop}%
\bibitem [{\citenamefont {Heinrich}\ \emph {et~al.}(2011)\citenamefont
  {Heinrich}, \citenamefont {Ludwig}, \citenamefont {Qian}, \citenamefont
  {Kubala},\ and\ \citenamefont {Marquardt}}]{heinrich_collective_2011}%
  \BibitemOpen
  \bibfield  {author} {\bibinfo {author} {\bibfnamefont {G.}~\bibnamefont
  {Heinrich}}, \bibinfo {author} {\bibfnamefont {M.}~\bibnamefont {Ludwig}},
  \bibinfo {author} {\bibfnamefont {J.}~\bibnamefont {Qian}}, \bibinfo {author}
  {\bibfnamefont {B.}~\bibnamefont {Kubala}}, \ and\ \bibinfo {author}
  {\bibfnamefont {F.}~\bibnamefont {Marquardt}},\ }\href {\doibase
  10.1103/PhysRevLett.107.043603} {\bibfield  {journal} {\bibinfo  {journal}
  {Physical Review Letters}\ }\textbf {\bibinfo {volume} {107}},\ \bibinfo
  {pages} {043603} (\bibinfo {year} {2011})}\BibitemShut {NoStop}%
\bibitem [{\citenamefont {Holmes}\ \emph {et~al.}(2012)\citenamefont {Holmes},
  \citenamefont {Meaney},\ and\ \citenamefont
  {Milburn}}]{holmes_synchronization_2012}%
  \BibitemOpen
  \bibfield  {author} {\bibinfo {author} {\bibfnamefont {C.~A.}\ \bibnamefont
  {Holmes}}, \bibinfo {author} {\bibfnamefont {C.~P.}\ \bibnamefont {Meaney}},
  \ and\ \bibinfo {author} {\bibfnamefont {G.~J.}\ \bibnamefont {Milburn}},\
  }\href {\doibase 10.1103/PhysRevE.85.066203} {\bibfield  {journal} {\bibinfo
  {journal} {Physical Review E}\ }\textbf {\bibinfo {volume} {85}},\ \bibinfo
  {pages} {066203} (\bibinfo {year} {2012})}\BibitemShut {NoStop}%
\bibitem [{\citenamefont {Xuereb}\ \emph {et~al.}(2012)\citenamefont {Xuereb},
  \citenamefont {Genes},\ and\ \citenamefont {Dantan}}]{xuereb_strong_2012}%
  \BibitemOpen
  \bibfield  {author} {\bibinfo {author} {\bibfnamefont {A.}~\bibnamefont
  {Xuereb}}, \bibinfo {author} {\bibfnamefont {C.}~\bibnamefont {Genes}}, \
  and\ \bibinfo {author} {\bibfnamefont {A.}~\bibnamefont {Dantan}},\ }\href
  {\doibase 10.1103/PhysRevLett.109.223601} {\bibfield  {journal} {\bibinfo
  {journal} {Physical Review Letters}\ }\textbf {\bibinfo {volume} {109}},\
  \bibinfo {pages} {223601} (\bibinfo {year} {2012})}\BibitemShut {NoStop}%
\bibitem [{\citenamefont {Schmidt}\ \emph {et~al.}(2012)\citenamefont
  {Schmidt}, \citenamefont {Ludwig},\ and\ \citenamefont
  {Marquardt}}]{schmidt_optomechanical_2012}%
  \BibitemOpen
  \bibfield  {author} {\bibinfo {author} {\bibfnamefont {M.}~\bibnamefont
  {Schmidt}}, \bibinfo {author} {\bibfnamefont {M.}~\bibnamefont {Ludwig}}, \
  and\ \bibinfo {author} {\bibfnamefont {F.}~\bibnamefont {Marquardt}},\ }\href
  {\doibase 10.1088/1367-2630/14/12/125005} {\bibfield  {journal} {\bibinfo
  {journal} {New Journal of Physics}\ }\textbf {\bibinfo {volume} {14}},\
  \bibinfo {pages} {125005} (\bibinfo {year} {2012})}\BibitemShut {NoStop}%
\bibitem [{\citenamefont {Ludwig}\ and\ \citenamefont
  {Marquardt}(2013)}]{ludwig_quantum_2013}%
  \BibitemOpen
  \bibfield  {author} {\bibinfo {author} {\bibfnamefont {M.}~\bibnamefont
  {Ludwig}}\ and\ \bibinfo {author} {\bibfnamefont {F.}~\bibnamefont
  {Marquardt}},\ }\href {\doibase 10.1103/PhysRevLett.111.073603} {\bibfield
  {journal} {\bibinfo  {journal} {Physical Review Letters}\ }\textbf {\bibinfo
  {volume} {111}},\ \bibinfo {pages} {073603} (\bibinfo {year}
  {2013})}\BibitemShut {NoStop}%
\bibitem [{\citenamefont {Xuereb}\ \emph {et~al.}(2014)\citenamefont {Xuereb},
  \citenamefont {Genes}, \citenamefont {Pupillo}, \citenamefont {Paternostro},\
  and\ \citenamefont {Dantan}}]{xuereb_reconfigurable_2014}%
  \BibitemOpen
  \bibfield  {author} {\bibinfo {author} {\bibfnamefont {A.}~\bibnamefont
  {Xuereb}}, \bibinfo {author} {\bibfnamefont {C.}~\bibnamefont {Genes}},
  \bibinfo {author} {\bibfnamefont {G.}~\bibnamefont {Pupillo}}, \bibinfo
  {author} {\bibfnamefont {M.}~\bibnamefont {Paternostro}}, \ and\ \bibinfo
  {author} {\bibfnamefont {A.}~\bibnamefont {Dantan}},\ }\href {\doibase
  10.1103/PhysRevLett.112.133604} {\bibfield  {journal} {\bibinfo  {journal}
  {Physical Review Letters}\ }\textbf {\bibinfo {volume} {112}},\ \bibinfo
  {pages} {133604} (\bibinfo {year} {2014})}\BibitemShut {NoStop}%
\bibitem [{\citenamefont {Lauter}\ \emph {et~al.}(2015)\citenamefont {Lauter},
  \citenamefont {Brendel}, \citenamefont {Habraken},\ and\ \citenamefont
  {Marquardt}}]{lauter_pattern_2015}%
  \BibitemOpen
  \bibfield  {author} {\bibinfo {author} {\bibfnamefont {R.}~\bibnamefont
  {Lauter}}, \bibinfo {author} {\bibfnamefont {C.}~\bibnamefont {Brendel}},
  \bibinfo {author} {\bibfnamefont {S.~J.~M.}\ \bibnamefont {Habraken}}, \ and\
  \bibinfo {author} {\bibfnamefont {F.}~\bibnamefont {Marquardt}},\ }\href
  {\doibase 10.1103/PhysRevE.92.012902} {\bibfield  {journal} {\bibinfo
  {journal} {Physical Review E}\ }\textbf {\bibinfo {volume} {92}},\ \bibinfo
  {pages} {012902} (\bibinfo {year} {2015})}\BibitemShut {NoStop}%
\bibitem [{\citenamefont {Schmidt}\ \emph
  {et~al.}(2015{\natexlab{a}})\citenamefont {Schmidt}, \citenamefont {Kessler},
  \citenamefont {Peano}, \citenamefont {Painter},\ and\ \citenamefont
  {Marquardt}}]{schmidt_optomechanical_2015}%
  \BibitemOpen
  \bibfield  {author} {\bibinfo {author} {\bibfnamefont {M.}~\bibnamefont
  {Schmidt}}, \bibinfo {author} {\bibfnamefont {S.}~\bibnamefont {Kessler}},
  \bibinfo {author} {\bibfnamefont {V.}~\bibnamefont {Peano}}, \bibinfo
  {author} {\bibfnamefont {O.}~\bibnamefont {Painter}}, \ and\ \bibinfo
  {author} {\bibfnamefont {F.}~\bibnamefont {Marquardt}},\ }\href {\doibase
  10.1364/OPTICA.2.000635} {\bibfield  {journal} {\bibinfo  {journal} {Optica}\
  }\textbf {\bibinfo {volume} {2}},\ \bibinfo {pages} {635} (\bibinfo {year}
  {2015}{\natexlab{a}})}\BibitemShut {NoStop}%
\bibitem [{\citenamefont {Schmidt}\ \emph
  {et~al.}(2015{\natexlab{b}})\citenamefont {Schmidt}, \citenamefont {Peano},\
  and\ \citenamefont {Marquardt}}]{schmidt_optomechanical_2015-1}%
  \BibitemOpen
  \bibfield  {author} {\bibinfo {author} {\bibfnamefont {M.}~\bibnamefont
  {Schmidt}}, \bibinfo {author} {\bibfnamefont {V.}~\bibnamefont {Peano}}, \
  and\ \bibinfo {author} {\bibfnamefont {F.}~\bibnamefont {Marquardt}},\ }\href
  {\doibase 10.1088/1367-2630/17/2/023025} {\bibfield  {journal} {\bibinfo
  {journal} {New Journal of Physics}\ }\textbf {\bibinfo {volume} {17}},\
  \bibinfo {pages} {023025} (\bibinfo {year} {2015}{\natexlab{b}})}\BibitemShut
  {NoStop}%
\bibitem [{\citenamefont {Zhang}\ \emph {et~al.}(2015)\citenamefont {Zhang},
  \citenamefont {Shah}, \citenamefont {Cardenas},\ and\ \citenamefont
  {Lipson}}]{zhang_synchronization_2015}%
  \BibitemOpen
  \bibfield  {author} {\bibinfo {author} {\bibfnamefont {M.}~\bibnamefont
  {Zhang}}, \bibinfo {author} {\bibfnamefont {S.}~\bibnamefont {Shah}},
  \bibinfo {author} {\bibfnamefont {J.}~\bibnamefont {Cardenas}}, \ and\
  \bibinfo {author} {\bibfnamefont {M.}~\bibnamefont {Lipson}},\ }\href
  {\doibase 10.1103/PhysRevLett.115.163902} {\bibfield  {journal} {\bibinfo
  {journal} {Physical Review Letters}\ }\textbf {\bibinfo {volume} {115}},\
  \bibinfo {pages} {163902} (\bibinfo {year} {2015})}\BibitemShut {NoStop}%
\bibitem [{\citenamefont {Peano}\ \emph {et~al.}(2015)\citenamefont {Peano},
  \citenamefont {Brendel}, \citenamefont {Schmidt},\ and\ \citenamefont
  {Marquardt}}]{peano_topological_2015}%
  \BibitemOpen
  \bibfield  {author} {\bibinfo {author} {\bibfnamefont {V.}~\bibnamefont
  {Peano}}, \bibinfo {author} {\bibfnamefont {C.}~\bibnamefont {Brendel}},
  \bibinfo {author} {\bibfnamefont {M.}~\bibnamefont {Schmidt}}, \ and\
  \bibinfo {author} {\bibfnamefont {F.}~\bibnamefont {Marquardt}},\ }\href
  {\doibase 10.1103/PhysRevX.5.031011} {\bibfield  {journal} {\bibinfo
  {journal} {Phys. Rev. X}\ }\textbf {\bibinfo {volume} {5}},\ \bibinfo {pages}
  {031011} (\bibinfo {year} {2015})}\BibitemShut {NoStop}%
\bibitem [{\citenamefont {Gan}\ \emph {et~al.}(2016)\citenamefont {Gan},
  \citenamefont {Xiong}, \citenamefont {Si}, \citenamefont {Lü},\ and\
  \citenamefont {Wu}}]{gan_solitons_2016}%
  \BibitemOpen
  \bibfield  {author} {\bibinfo {author} {\bibfnamefont {J.-H.}\ \bibnamefont
  {Gan}}, \bibinfo {author} {\bibfnamefont {H.}~\bibnamefont {Xiong}}, \bibinfo
  {author} {\bibfnamefont {L.-G.}\ \bibnamefont {Si}}, \bibinfo {author}
  {\bibfnamefont {X.-Y.}\ \bibnamefont {Lü}}, \ and\ \bibinfo {author}
  {\bibfnamefont {Y.}~\bibnamefont {Wu}},\ }\href {\doibase
  10.1364/OL.41.002676} {\bibfield  {journal} {\bibinfo  {journal} {Optics
  Letters}\ }\textbf {\bibinfo {volume} {41}},\ \bibinfo {pages} {2676}
  (\bibinfo {year} {2016})}\BibitemShut {NoStop}%
\bibitem [{\citenamefont {Mathew}\ \emph {et~al.}()\citenamefont {Mathew},
  \citenamefont {del Pino},\ and\ \citenamefont
  {Verhagen}}]{mathew_synthetic_2018}%
  \BibitemOpen
  \bibfield  {author} {\bibinfo {author} {\bibfnamefont {J.~P.}\ \bibnamefont
  {Mathew}}, \bibinfo {author} {\bibfnamefont {J.}~\bibnamefont {del Pino}}, \
  and\ \bibinfo {author} {\bibfnamefont {E.}~\bibnamefont {Verhagen}},\
  }\href@noop {} {\ }\Eprint {http://arxiv.org/abs/1812.09369}
  {arXiv:1812.09369 [physics.optics]} \BibitemShut {NoStop}%
\bibitem [{\citenamefont {Hafezi}\ and\ \citenamefont
  {Rabl}(2012)}]{hafezi_optomechanically_2012}%
  \BibitemOpen
  \bibfield  {author} {\bibinfo {author} {\bibfnamefont {M.}~\bibnamefont
  {Hafezi}}\ and\ \bibinfo {author} {\bibfnamefont {P.}~\bibnamefont {Rabl}},\
  }\href {\doibase 10.1364/OE.20.007672} {\bibfield  {journal} {\bibinfo
  {journal} {Opt. Express}\ }\textbf {\bibinfo {volume} {20}},\ \bibinfo
  {pages} {7672} (\bibinfo {year} {2012})}\BibitemShut {NoStop}%
\bibitem [{\citenamefont {Hafezi}\ \emph {et~al.}(2011)\citenamefont {Hafezi},
  \citenamefont {Demler}, \citenamefont {Lukin},\ and\ \citenamefont
  {Taylor}}]{hafezi_robust_2011}%
  \BibitemOpen
  \bibfield  {author} {\bibinfo {author} {\bibfnamefont {M.}~\bibnamefont
  {Hafezi}}, \bibinfo {author} {\bibfnamefont {E.~A.}\ \bibnamefont {Demler}},
  \bibinfo {author} {\bibfnamefont {M.~D.}\ \bibnamefont {Lukin}}, \ and\
  \bibinfo {author} {\bibfnamefont {J.~M.}\ \bibnamefont {Taylor}},\ }\href
  {https://doi.org/10.1038/nphys2063} {\bibfield  {journal} {\bibinfo
  {journal} {Nature Physics}\ }\textbf {\bibinfo {volume} {7}},\ \bibinfo
  {pages} {907 EP } (\bibinfo {year} {2011})}\BibitemShut {NoStop}%
\bibitem [{\citenamefont {Hafezi}\ \emph {et~al.}(2013)\citenamefont {Hafezi},
  \citenamefont {Mittal}, \citenamefont {Fan}, \citenamefont {Migdall},\ and\
  \citenamefont {Taylor}}]{hafezi_imaging_2013}%
  \BibitemOpen
  \bibfield  {author} {\bibinfo {author} {\bibfnamefont {M.}~\bibnamefont
  {Hafezi}}, \bibinfo {author} {\bibfnamefont {S.}~\bibnamefont {Mittal}},
  \bibinfo {author} {\bibfnamefont {J.}~\bibnamefont {Fan}}, \bibinfo {author}
  {\bibfnamefont {A.}~\bibnamefont {Migdall}}, \ and\ \bibinfo {author}
  {\bibfnamefont {J.~M.}\ \bibnamefont {Taylor}},\ }\href {\doibase
  10.1038/nphoton.2013.274} {\bibfield  {journal} {\bibinfo  {journal} {Nature
  Photonics}\ }\textbf {\bibinfo {volume} {7}},\ \bibinfo {pages} {1001}
  (\bibinfo {year} {2013})}\BibitemShut {NoStop}%
\bibitem [{\citenamefont {Mousavi}\ \emph {et~al.}(2015)\citenamefont
  {Mousavi}, \citenamefont {Khanikaev},\ and\ \citenamefont
  {Wang}}]{mousavi_topologically_2015}%
  \BibitemOpen
  \bibfield  {author} {\bibinfo {author} {\bibfnamefont {S.~H.}\ \bibnamefont
  {Mousavi}}, \bibinfo {author} {\bibfnamefont {A.~B.}\ \bibnamefont
  {Khanikaev}}, \ and\ \bibinfo {author} {\bibfnamefont {Z.}~\bibnamefont
  {Wang}},\ }\href {\doibase 10.1038/ncomms9682} {\bibfield  {journal}
  {\bibinfo  {journal} {Nature Communications}\ }\textbf {\bibinfo {volume}
  {6}},\ \bibinfo {pages} {8682} (\bibinfo {year} {2015})}\BibitemShut
  {NoStop}%
\bibitem [{\citenamefont {Brendel}\ \emph {et~al.}(2017)\citenamefont
  {Brendel}, \citenamefont {Peano}, \citenamefont {Painter},\ and\
  \citenamefont {Marquardt}}]{brendel_pseudomagnetic_2017}%
  \BibitemOpen
  \bibfield  {author} {\bibinfo {author} {\bibfnamefont {C.}~\bibnamefont
  {Brendel}}, \bibinfo {author} {\bibfnamefont {V.}~\bibnamefont {Peano}},
  \bibinfo {author} {\bibfnamefont {O.~J.}\ \bibnamefont {Painter}}, \ and\
  \bibinfo {author} {\bibfnamefont {F.}~\bibnamefont {Marquardt}},\ }\href
  {https://www.pnas.org/content/114/17/E3390} {\bibfield  {journal} {\bibinfo
  {journal} {Proc. Nat. Acad. Sci. U.S.A.}\ }\textbf {\bibinfo {volume} {114}}
  (\bibinfo {year} {2017})}\BibitemShut {NoStop}%
\bibitem [{\citenamefont {Brendel}\ \emph {et~al.}(2018)\citenamefont
  {Brendel}, \citenamefont {Peano}, \citenamefont {Painter},\ and\
  \citenamefont {Marquardt}}]{brendel_snowflake_2018}%
  \BibitemOpen
  \bibfield  {author} {\bibinfo {author} {\bibfnamefont {C.}~\bibnamefont
  {Brendel}}, \bibinfo {author} {\bibfnamefont {V.}~\bibnamefont {Peano}},
  \bibinfo {author} {\bibfnamefont {O.}~\bibnamefont {Painter}}, \ and\
  \bibinfo {author} {\bibfnamefont {F.}~\bibnamefont {Marquardt}},\ }\href
  {\doibase 10.1103/PhysRevB.97.020102} {\bibfield  {journal} {\bibinfo
  {journal} {Physical Review B}\ }\textbf {\bibinfo {volume} {97}},\ \bibinfo
  {pages} {020102} (\bibinfo {year} {2018})}\BibitemShut {NoStop}%
\bibitem [{\citenamefont {Yu}\ \emph {et~al.}(2018)\citenamefont {Yu},
  \citenamefont {He}, \citenamefont {Wang}, \citenamefont {Liu}, \citenamefont
  {Sun}, \citenamefont {Li}, \citenamefont {Lu}, \citenamefont {Lu},
  \citenamefont {Liu},\ and\ \citenamefont {Chen}}]{yu_elastic_2018}%
  \BibitemOpen
  \bibfield  {author} {\bibinfo {author} {\bibfnamefont {S.-Y.}\ \bibnamefont
  {Yu}}, \bibinfo {author} {\bibfnamefont {C.}~\bibnamefont {He}}, \bibinfo
  {author} {\bibfnamefont {Z.}~\bibnamefont {Wang}}, \bibinfo {author}
  {\bibfnamefont {F.-K.}\ \bibnamefont {Liu}}, \bibinfo {author} {\bibfnamefont
  {X.-C.}\ \bibnamefont {Sun}}, \bibinfo {author} {\bibfnamefont
  {Z.}~\bibnamefont {Li}}, \bibinfo {author} {\bibfnamefont {H.-Z.}\
  \bibnamefont {Lu}}, \bibinfo {author} {\bibfnamefont {M.-H.}\ \bibnamefont
  {Lu}}, \bibinfo {author} {\bibfnamefont {X.-P.}\ \bibnamefont {Liu}}, \ and\
  \bibinfo {author} {\bibfnamefont {Y.-F.}\ \bibnamefont {Chen}},\ }\href
  {\doibase 10.1038/s41467-018-05461-5} {\bibfield  {journal} {\bibinfo
  {journal} {Nature Communications}\ }\textbf {\bibinfo {volume} {9}},\
  \bibinfo {pages} {3072} (\bibinfo {year} {2018})}\BibitemShut {NoStop}%
\bibitem [{\citenamefont {Cha}\ \emph {et~al.}(2018)\citenamefont {Cha},
  \citenamefont {Kim},\ and\ \citenamefont {Daraio}}]{cha_experimental_2018}%
  \BibitemOpen
  \bibfield  {author} {\bibinfo {author} {\bibfnamefont {J.}~\bibnamefont
  {Cha}}, \bibinfo {author} {\bibfnamefont {K.~W.}\ \bibnamefont {Kim}}, \ and\
  \bibinfo {author} {\bibfnamefont {C.}~\bibnamefont {Daraio}},\ }\href
  {\doibase 10.1038/s41586-018-0764-0} {\bibfield  {journal} {\bibinfo
  {journal} {Nature (London)}\ }\textbf {\bibinfo {volume} {564}},\ \bibinfo
  {pages} {229} (\bibinfo {year} {2018})}\BibitemShut {NoStop}%
\bibitem [{\citenamefont {Gardiner}\ and\ \citenamefont
  {Collett}(1985)}]{Gardiner_1985}%
  \BibitemOpen
  \bibfield  {author} {\bibinfo {author} {\bibfnamefont {C.~W.}\ \bibnamefont
  {Gardiner}}\ and\ \bibinfo {author} {\bibfnamefont {M.~J.}\ \bibnamefont
  {Collett}},\ }\href {\doibase 10.1103/PhysRevA.31.3761} {\bibfield  {journal}
  {\bibinfo  {journal} {Phys. Rev. A}\ }\textbf {\bibinfo {volume} {31}},\
  \bibinfo {pages} {3761} (\bibinfo {year} {1985})}\BibitemShut {NoStop}%
\bibitem [{\citenamefont {{Th{\`a}nh Nam}}\ \emph {et~al.}()\citenamefont
  {{Th{\`a}nh Nam}}, \citenamefont {{Napi{\'o}rkowski}},\ and\ \citenamefont
  {{Solovej}}}]{Nam_2015}%
  \BibitemOpen
  \bibfield  {author} {\bibinfo {author} {\bibfnamefont {P.}~\bibnamefont
  {{Th{\`a}nh Nam}}}, \bibinfo {author} {\bibfnamefont {M.}~\bibnamefont
  {{Napi{\'o}rkowski}}}, \ and\ \bibinfo {author} {\bibfnamefont {J.~P.}\
  \bibnamefont {{Solovej}}},\ }\href {http://arxiv.org/abs/1508.07321}
  {\bibinfo  {journal} {arXiv:1508.07321[math-ph]}\ }\BibitemShut {NoStop}%
\bibitem [{\citenamefont {Verhagen}\ \emph {et~al.}(2012)\citenamefont
  {Verhagen}, \citenamefont {Del{\'e}glise}, \citenamefont {Weis},
  \citenamefont {Schliesser},\ and\ \citenamefont {Kippenberg}}]{Verhagen2012}%
  \BibitemOpen
\bibfield  {journal} {  }\bibfield  {author} {\bibinfo {author} {\bibfnamefont
  {E.}~\bibnamefont {Verhagen}}, \bibinfo {author} {\bibfnamefont
  {S.}~\bibnamefont {Del{\'e}glise}}, \bibinfo {author} {\bibfnamefont
  {S.}~\bibnamefont {Weis}}, \bibinfo {author} {\bibfnamefont {A.}~\bibnamefont
  {Schliesser}}, \ and\ \bibinfo {author} {\bibfnamefont {T.~J.}\ \bibnamefont
  {Kippenberg}},\ }\href {https://doi.org/10.1038/nature10787} {\bibfield
  {journal} {\bibinfo  {journal} {Nature (London)}\ }\textbf {\bibinfo {volume}
  {482}},\ \bibinfo {pages} {63 EP } (\bibinfo {year} {2012})}\BibitemShut
  {NoStop}%
\end{thebibliography}%

\end{document}